\documentclass[3p,times,preprint]{elsarticle}
\usepackage{ecrc}

\volume{00}

\firstpage{1}

\journalname{Procedia Engineering}

\runauth{J. Marcon et al.}

\jid{proeng}

\usepackage{amssymb}

 \biboptions{sort&compress}

\usepackage[figuresright]{rotating}

\usepackage{color}
\usepackage{tikz}
\usepackage{bm}
\usepackage{subcaption}
\usepackage{amsmath}

\begin{document}

\begin{frontmatter}



\dochead{26th International Meshing Roundtable, IMR26, 18--21 September 2017, Barcelona, Spain}

\title{A variational approach to high-order \textit{r}-adaptation}


\author[a]{Julian Marcon}
\author[a]{Michael Turner}
\author[b]{David Moxey}
\author[a]{Spencer J. Sherwin}
\author[a]{Joaquim Peir\'{o}\corref{cor1}}

\address[a]{Department of Aeronautics, Imperial College London, South Kensington Campus, London SW7 2AZ, United Kingdom}
\address[b]{College of Engineering, Mathematics and Physical Sciences, University of Exeter, Exeter EX4 4QF, United Kingdom}

\begin{abstract}

A variational framework, initially developed for high-order mesh optimisation, is being extended for \textit{r}-adaptation.
The method is based on the minimisation of a functional of the mesh deformation.
To achieve adaptation, elements of the initial mesh are manipulated using metric tensors to obtain target elements.
The nonlinear optimisation in turns adapts the final high-order mesh to best fit the description of the target elements by minimising the element distortion.
Encouraging preliminary results prove that the method behaves well and can be used in the future for more extensive work which shall include the use of error indicators from CFD simulations.

\end{abstract}

\begin{keyword}
high-order meshing \sep{} \textit{r}-adaptation \sep{} variational formulation




\end{keyword}
\cortext[cor1]{Corresponding author.}
\end{frontmatter}





\section{Introduction}\label{sec:intro}

High-order computational fluid dynamics (CFD) methods, such as the spectral/\textit{hp} element methods, are able to capture a large range of both temporal and spatial scales thanks to their low dispersion and diffusion error~\cite{karniadakis-2005}.
They are however highly susceptible to inaccuracies in the geometrical representation of computational domains.
For that reason, curvilinear high-order meshes must be used to attain the expected exponential rates of convergence.
One typical procedure used to generate curvilinear high-order meshes involves the transformation of a coarse linear mesh into a geometry-accurate high-order mesh by projection of high-order nodes onto the boundaries.
This procedure is unfortunately prone to the creation of highly deformed and sometimes invalid elements and optimisation of the mesh is often required.

Recent work has been made by Turner et al.~\cite{Turner2016a} to unify past approaches to mesh optimisation into a generalised framework.
A variational approach is used in which a functional of the deformation energy is optimised by a nonlinear algorithm.
The approach has proved to be robust in its ability to improve meshes and turned out to be highly scalable.
The optimisation process is applied on the mapping between the ideal and the curvilinear high-order elements.
This directly results in the relocation of nodes to minimise the deformation energy with respect to the ideal element.

The purpose of this work is to demonstrate the extensibility of this method to mesh adaptation.
Adaptive meshes most often rely on a number of mesh manipulation strategies such as local or global remeshing, edge manipulation, element splitting/collapsing or a combination of them.
This research note however deals exclusively with moving meshes where nodes are relocated, also known as \textit{r}-adaptation.
\textit{r}-adaptation has been studied for many years and it has interesting advantages~\cite{Huang+Russell-2011, Budd2009} in comparison with the other two types of adaptation, \textit{h}- and \textit{p}-.
Firstly it keeps the number of degrees of freedom constant which is important when computational resources are limited.
For a given amount of resources, one would be able to refine in regions of interest by relocating nodes, and therefore degrees of freedom, in said regions.
This implies a loss of accuracy in coarsened regions, which is however insignificant in comparison to the global gain of accuracy.
Additionally, moving meshes preserve connectivities and therefore keep data structures unchanged.
The construction of data structures in a distributed memory parallel context is always accompanied by large overhead.
In this context, \textit{r}-adaptation could prove useful in unsteady simulations for flow feature tracking without the need of a restart after every adaptation of the mesh.

This variational approach to \textit{r}-adaptation proposes to use the algorithm employed in variational mesh optimisation to achieve refinement in target areas.
While the mesh optimisation procedure operates on the linear-to-curvilinear mapping, adaptation naturally occurs when the mapping between the reference and the linear elements is modified by means of metric tensors.
Section~\ref{sec:var} provides an overview of the variational framework with Section~\ref{sec:opti} summarising the approach to mesh optimisation and Section~\ref{sec:r-adapt} describing the transformation of the framework for \textit{r}-adaptation.
In Section~\ref{sec:res}, we present preliminary results of adaptive meshes using analytical target functions.
We bring this research note to an end in Section~\ref{sec:concl} with a quick summary and an outlook to future developments and challenges. 

\section{Variational framework}\label{sec:var}

\subsection{Mesh optimisation}\label{sec:opti}

\begin{figure}
  \centering
  \includegraphics[width=0.6\textwidth]{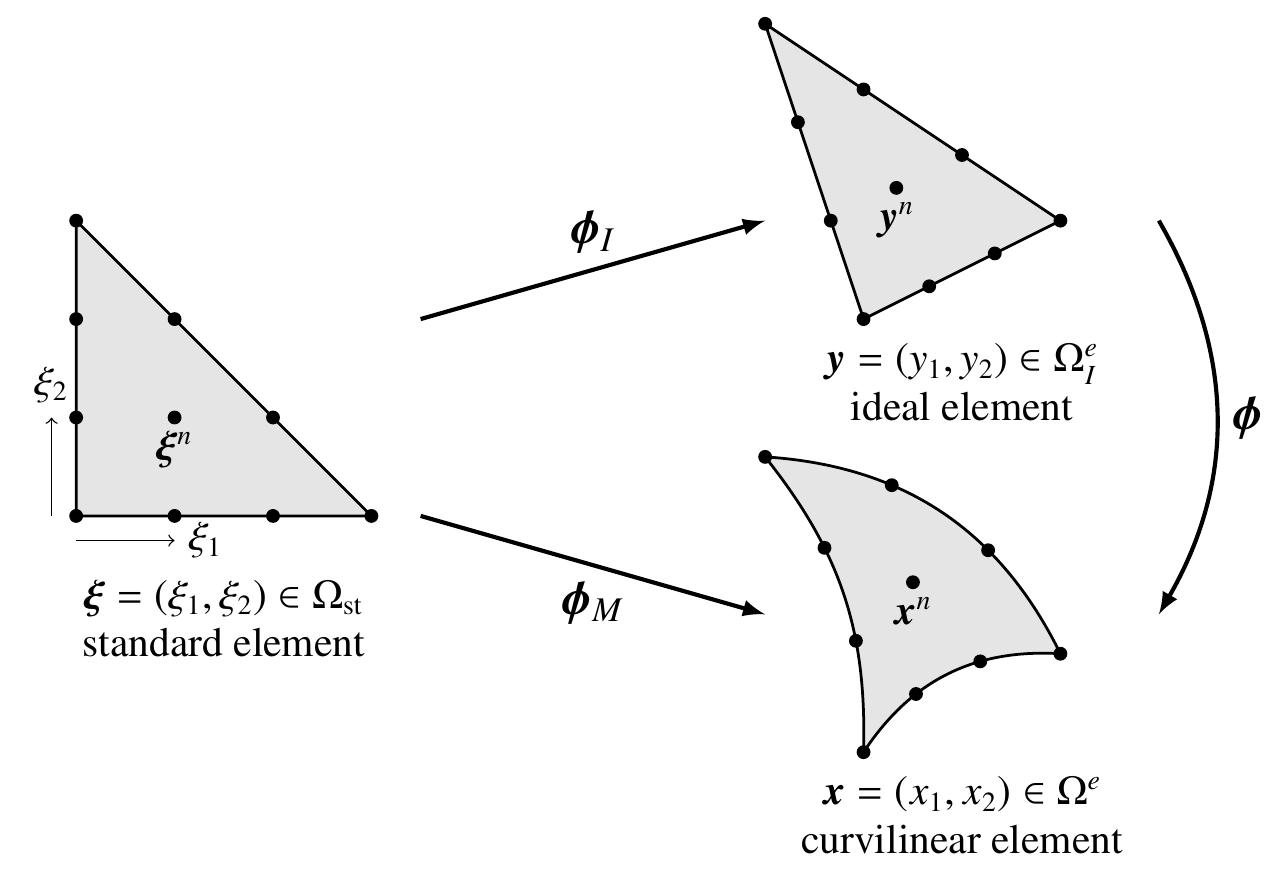}
  \caption{Existing mappings between the reference, the ideal and the curvilinear elements.}\label{fig:mapping}
\end{figure}

The variational framework used in these developments is based on the work of Turner et al.~\cite{Turner2016a}.
A curvilinear high-order element \( \bm{\Omega}^e \) can be represented by a mapping \( \bm{\phi}_M \) of a reference element \( \bm{\Omega}_{st} \) as shown in Fig.~\ref{fig:mapping}.
Mapping \( \bm{\phi}_M \) can in turn be decomposed into a reference-to-ideal mapping \( \bm{\phi}_I \) and an ideal-to-curvilinear mapping \( \bm{\phi} \).
The high-order linear intermediate element \( \bm{\Omega}_I^e \) is an ideal element from which deformation energy is computed.

The mesh is deformed to minimise an energy functional \( \mathcal{E} (\nabla \bm{\phi}) \), a function of the mesh deformation \( \bm{\phi} \):
\[
  \mathrm{find} \, \min_{\bm{\phi}} \mathcal{E} (\nabla \bm{\phi}) =
    \int_{\bm{\Omega}^e} W(\nabla \bm{\phi}) d\bm{y}
\]

The function \( \mathcal{E} (\nabla \bm{\phi}) \) can take different forms depending on the formulation of the deformation energy.
In this work, although other formulations could also have been used, the hyperelastic elastic model~\cite{Persson+Peraire-2009} has been retained for its demonstrated efficiency in previous works~\cite{Turner2016a}. In the variational framework, the hyperelastic strain energy takes the form of
\[
  W = \frac{\mu}{2} (I_1^{\bm{C}} - 3) - \mu \ln~J + \frac{\lambda}{2} {(\ln~J)}^2
\]
where \( \lambda \) and \( \mu \) are material constants, \( \bm{C} \) is the right Cauchy-Green tensor, \( I_1^{\bm{C}} \) is its trace and \( J \) is the determinant of the Jacobian matrix \( \bm{J} = \nabla \bm{\phi} \).

\subsection{\textit{r}-adaptation}\label{sec:r-adapt}

In the framework of adaptive meshes, the ideal element \( \bm{\Omega}_I^e \) becomes a target element \( \bm{\Omega}_T^e \).
Assuming that this target element \( \bm{\Omega}_T^e \) is modified, the optimisation of the mesh by minimisation of the energy functional \( \mathcal{E} (\nabla \bm{\phi}) \) will force element \( \bm{\Omega}^e \) towards a shape and dimensions similar to \( \bm{\Omega}_T^e \).
The manipulation of \( \bm{\Omega}_T^e \) is achieved by modification of its mapping \( \bm{\phi}_T \) and can be isotropic or anisotropic alike.
Linear transformations can be applied to the Jacobian of the mapping \( \bm{J}^T = \nabla \bm{\phi}^T \).
In the anisotropic case, the Jacobian is multiplied by a metric tensor \( \bm{M} \):
\[
\bm{J}^T = \bm{M} \bm{J}^I
\]
In the isotropic case, the Jacobian is simply scaled by a linear factor \( r \), which can in turn be expressed more generally as a metric tensor \( r \bm{I} \) multiplication:
\[
\bm{J}^T = r \bm{J}^I = (r \bm{I}) \bm{J}^I
\]

The developments hereby presented were implemented in \textbf{NekMesh}, an open-source softare solution for the generation of geometry-accurate high-order meshes, part of the \textbf{Nektar++} platform~\cite{Cantwell2015}.
Section~\ref{sec:res} presents some preliminary results of adaptive meshes.

\section{Results}\label{sec:res}

\begin{figure}
  \begin{subfigure}{.5\textwidth}
    \centering
    \includegraphics[width=.8\linewidth]{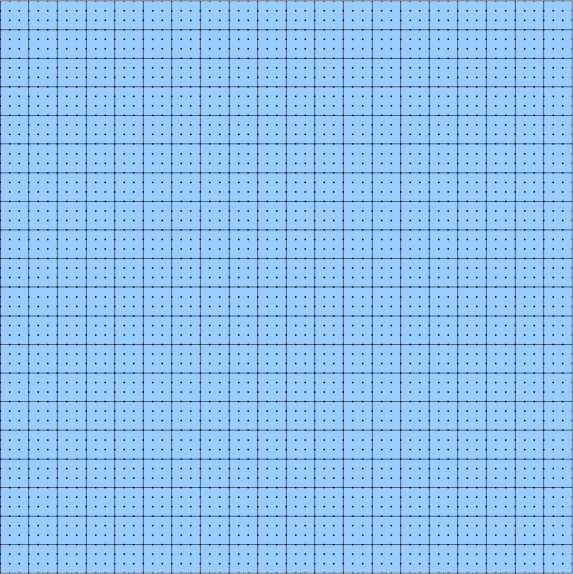}
    \caption{Initial mesh}\label{fig:initial-quad}
  \end{subfigure}
  \begin{subfigure}{.5\textwidth}
    \centering
    \includegraphics[width=.8\linewidth]{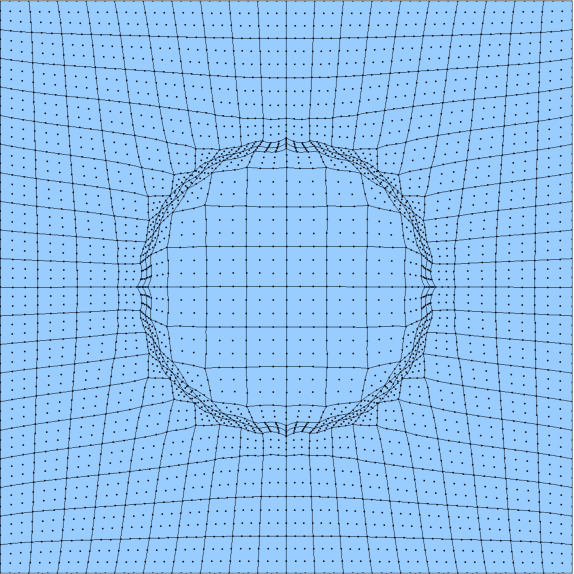}
    \caption{Adapted mesh}\label{fig:adapted-quad}
  \end{subfigure}
  \begin{subfigure}{\textwidth}
    \centering
    \includegraphics[width=0.8\linewidth]{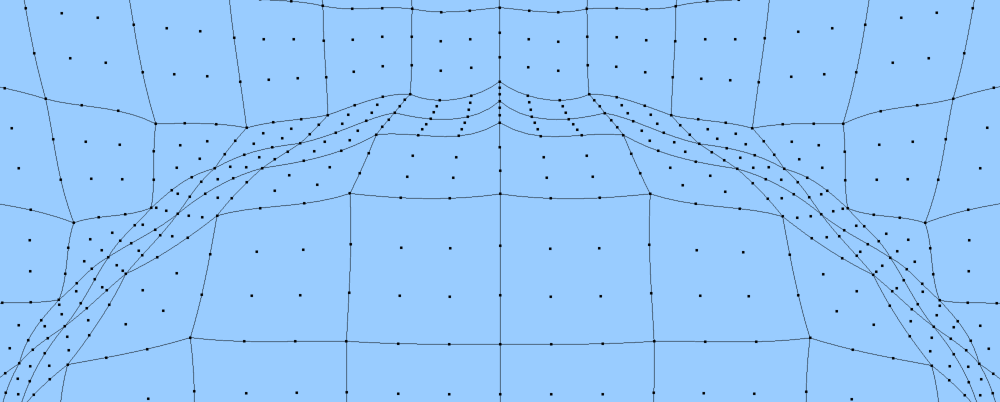}
    \caption{Zoom in on adapted mesh}\label{fig:zoom-quad}
  \end{subfigure}
  \caption{Anisotropic adaptation of a unit side quad mesh of constant 3\textsuperscript{rd} polynomial order along a circle circumference}\label{fig:adaptation-quad}
\end{figure}

Ideally, adaptation should be driven by an error indicator and this shall be the focus of future work.
In the scope of this research note, the feasibility of this method shall be demonstrated by using analytical expressions for the metric tensor.
The example hereby presented proposes to adapt a homogeneously meshed unit side domain such as the one shown in Fig.~\ref{fig:initial-quad}.
We aim at refining along the circumference of a circle of unit diameter.
This is achieved anisotropically by shrinking elements in the radial direction only.

A scaling factor \( r \) is defined in the radial direction at an angle \( \alpha \) from the x-axis.
The metric tensor can be expressed as a succession of linear 2D transformations:
\begin{enumerate}
  \item Rotate the element so that the radial axis coincides with the x-axis
    \[
      \bm{M}_1 =
        \begin{pmatrix}
          \cos \alpha  && \sin \alpha \\
          -\sin \alpha && \cos \alpha
        \end{pmatrix}
    \]
  \item Scale the element along the x-axis
    \[
      \bm{M}_2 =
        \begin{pmatrix}
          r && 0 \\
          0 && 1
        \end{pmatrix}
    \]
  \item Rotate the element back to its initial orientation
    \[
      \bm{M}_3 =
        \begin{pmatrix}
          \cos \alpha && -\sin \alpha \\
          \sin \alpha && \cos \alpha
        \end{pmatrix}
    \]
\end{enumerate}

The combined metric tensor becomes:
\[
  \bm{M} = \bm{M}_1 \bm{M}_2 \bm{M}_3 =
    \begin{pmatrix}
      1 + (r - 1) \cos^2 \alpha       && (r - 1) \sin \alpha \cos \alpha \\
      (r - 1) \sin \alpha \cos \alpha && 1 + (r - 1) \sin^2 \alpha
    \end{pmatrix}
\]

Additionally, a distribution of \( r(d) \) is defined, with \( d = \sqrt{x^2 + y^2} \) the distance from the centre of the domain, using a Gaussian distribution such as:
\[
  r(d) = 1 - \frac{A}{\sqrt{2 \pi \sigma^2}} e^{-\frac{ {(x - \mu)}^2 }{2 \sigma^2}}
\]
with the mean \( \mu = 0.5 \), the standard deviation \( \sigma = 0.05 \) and \( A = 0.9 \sqrt{2 \pi \sigma^2} \) such that \( \min_{d} r(d) = 0.1 \).

Results are shown for a quad mesh in Fig.~\ref{fig:adaptation-quad}.
The adapted mesh in Fig.~\ref{fig:adapted-quad} shows excellent refinement in the unit diameter circumference area.
Coarsening is also observed everwhere else with bigger elements noted inside the circle.
Such coarsening is to be expected as nodes are moved towards the unit diameter circumference and therefore stretch elements in the rest of the domain.
It can also be observed from Fig.~\ref{fig:zoom-quad} that adaptation is indeed anisotropic: elements are shrunk in the radial direction only, keeping the size in the angular direction constant.

\begin{figure}
  \begin{subfigure}{.5\textwidth}
    \centering
    \includegraphics[width=.8\linewidth]{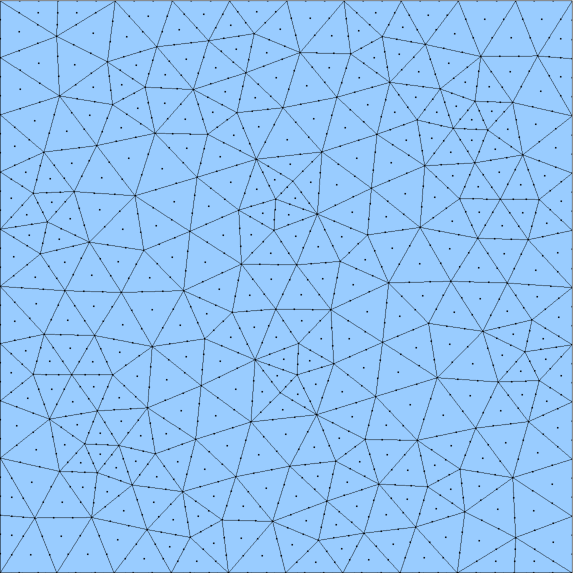}
    \caption{Initial mesh}\label{fig:initial-tri}
  \end{subfigure}
  \begin{subfigure}{.5\textwidth}
    \centering
    \includegraphics[width=.8\linewidth]{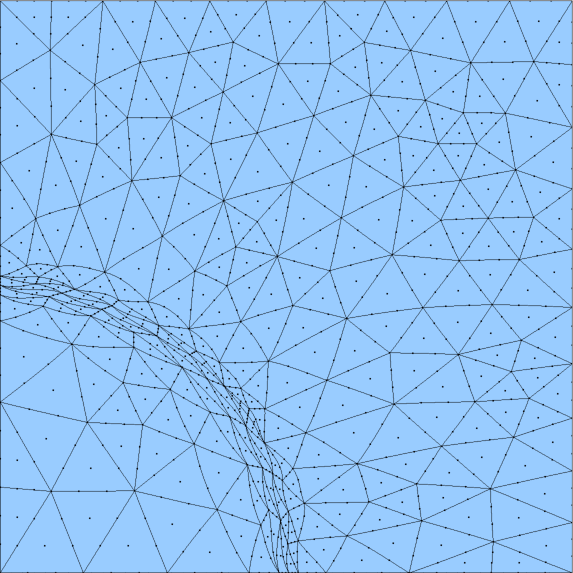}
    \caption{Adapted mesh}\label{fig:adapted-tri}
  \end{subfigure}
  \caption{Anisotropic adaptation of a triangular mesh of constant 3\textsuperscript{rd} polynomial order along a circle arc}\label{fig:adaptation-tri}
\end{figure}

Another example is shown in Fig.~\ref{fig:adaptation-tri}.
This domain corresponds to the truncated upper-right quadrant of the previous example, meshed this time by triangulation.
The method behaves equally well for a less uniformly distributed triangular mesh as shown in Fig.~\ref{fig:adapted-tri}.
The reader should note that this is a smaller domain, not a zoom in on a bigger domain, therefore demonstrating the CAD-sliding node capabilities of \textbf{NekMesh}.

In both examples, highly curved elements can be observed in the vicinity of the unit diameter region, which is indeed the expected behaviour.
At the present time, a single metric tensor is used per element, computed at the barycentre of the linear element.
This results in linear target elements, which are in turn deformed curvilinearly by the algorithm in ways not intended by the metric tensor.
Future developments shall attempt to produce curvilinear target elements by defining spatially varying metric tensors inside each element.

\section{Conclusions}\label{sec:concl}

We have presented a novel approach to \textit{r}-adaptation based on a variational approach for high-order meshes.
The algorithm, being based on the variational framework for mesh optimisation, retains all properties of the latter while extending its capabilities to adaptive high-order meshes.
The core of the adaptation relies on the manipulation of the ideal target elements by use of metric tensors.
First results were presented to demonstrate the feasibility of the method and establish a proof of concept for future work.
Such future work shall attempt to integrate error indicators in the adaptation process.
The challenge will lie in not only identifying a reliable error indicator but also converting this error indicator into a smoothly varying metric field, whether it be for isotropic or anisotropic adaptation.
				
\section*{Acknowledgements}

JM acknowledges funding from the European Union's Horizon 2020 research and innovation programme under the Marie Sk\l{}odowska-Curie grant agreement No 675008.
MT acknowledges funding from Airbus and EPSRC under an industrial CASE studentship.
DM acknowledges support from the EU Horizon 2020 project ExaFLOW (grant ID 671571) and the PRISM project under EPSRC grant EP/L000407/1.



\bibliography{Refs,Poster2017}
\bibliographystyle{model1-num-names}

\end{document}